\documentclass[pra,print,superscriptaddress,twocolumn]{revtex4}
\usepackage{amssymb}
\usepackage{amsmath}
\usepackage{epsfig}
\usepackage{color}
\usepackage{graphics, graphicx}
\usepackage{bbold}
\usepackage{psfrag}
\usepackage{mathcomp}
\usepackage{subfigure}
\usepackage{verbatim}
\usepackage{color}
\usepackage[colorlinks,citecolor=blue]{hyperref}

\begin{document}

\title{Dynamical Zeeman resonance in spin-orbit-coupled spin-1 Bose gases}
\author{Jingtao Fan}
\affiliation{State Key Laboratory of Quantum Optics and Quantum Optics Devices, Institute
of Laser Spectroscopy, Shanxi University, Taiyuan 030006, China}
\affiliation{Collaborative Innovation Center of Extreme Optics, Shanxi University,
Taiyuan 030006, China}
\author{Gang Chen}
\thanks{chengang971@163.com}
\affiliation{State Key Laboratory of Quantum Optics and Quantum Optics Devices, Institute
of Laser Spectroscopy, Shanxi University, Taiyuan 030006, China}
\affiliation{Collaborative Innovation Center of Extreme Optics, Shanxi University,
Taiyuan 030006, China}
\affiliation{Collaborative Innovation Center of Light Manipulations and Applications,
Shandong Normal University, Jinan 250358, China}
\author{Suotang Jia}
\affiliation{State Key Laboratory of Quantum Optics and Quantum Optics Devices, Institute
of Laser Spectroscopy, Shanxi University, Taiyuan 030006, China}
\affiliation{Collaborative Innovation Center of Extreme Optics, Shanxi University,
Taiyuan 030006, China}

\begin{abstract}
We predict a dynamical resonant effect, which is driven by externally
applied linear and quadratic Zeeman fields, in a spin-orbit-coupled spin-1
Bose-Einstein condensate. The Bose-Einstein condensate is assumed to be
initialized in some superposed state of Zeeman sublevels and subject to a
sudden shift of the trapping potential. It is shown that the time-averaged
center-of-mass oscillation and the spin polarizations of the Bose-Einstein
condensate exhibit remarkable resonant peaks when the Zeeman fields are
tuned to certain strengths. The underlying physics behind this resonance can
be traced back to the out-of-phase interference of the dynamical phases
carried by different spin-orbit states. By analyzing the single particle
spectrum, the resonant condition is summarized as a simple algebraic
relation, connecting the strengths of the linear and quadratic Zeeman
fields. This property is potentially applicable in quantum information and
quantum precision measurement.
\end{abstract}

\pacs{42.50.Pq}
\maketitle

\section{Introduction}

The impacts of gauge fields on quantum matters have been a central research
topic for lots of areas of physics, ranging from statistical mechanics \cite%
{Stat1,Stat2}, condensed-matter physics \cite{Berry,Topo1,Topo2}, to atomic
physics \cite{GaugeAtom1,GaugeAtom2}, etc. Among various forms of gauge
fields, the spin-orbit (SO) coupling is of particular interest as it is
naturally owned by electrons in solids and responsible for vast fundamental
physics such as topological insulators and superconductors \cite{Topo1,Topo2}%
. However, to some extend, a deep understanding of the SO-coupling-related
physics is hindered by the impurities and uncontrolled parameters in solid
state materials. In this context, ultracold atoms with synthetic SO coupling
have received much attention in resent years \cite{ASO1,ASO2,ASO3}. Not only
because it provides a versatile platform to simulate various novel quantum
phases, with precisely controllable parameters setting \cite%
{ASOP2,ASOP3,ASOP4,ASOP5,ASOP6,ASOP7,ASOP8,ASOP9,ASOP10,ASOP11,ASOP12,ASOP13,ASOP14,ASOP15}%
, but also due to its ability to engineer the interplay between spin and
orbit dynamics \cite{ASOD1,ASOD2,ASOD3,ASOD4,ASOD5,ASOD6}, which is of
potential usage for applications in atomtronics and spintronics.

While electrons moving in solids are intrinsically spin-half systems,
neutral atoms with rich hyperfine states could have higher spins, from which
one can construct not only the rank-1 spin vector, but also the rank-2
spin-quadruple tensor \cite{Hspin1,Hspin2,Hspin3,Hspin4,Hspin5,Hspin6,Hspin7}%
. This greatly enriches the SO-coupling-related physics emerging from the
spinor character of high spin systems \cite%
{SpinT1,SpinT2,SpinT3,SpinT4,SpinT5,SpinT6,SpinT7,SpinT8,SpinT9}. Indeed,
the SO coupling for spin-1 Bose-Einstein condensates (BECs) has been
experimentally realized through Raman coupling among three hyperfine states
\cite{SpinE1} or with the use of a gradient magnetic field \cite{SpinE2}.
Theoretical interest in this field is also tremendous. Notable examples
include the prediction of competing spin and nematic orders \cite{SpinT2},
multi-roton structures \cite{SpinT3}, and quantum multicriticalities \cite%
{SpinT4}. Recently, by coupling the rank-2 spin tensor to linear \cite%
{SpinT6,SpinT7} or orbital angular momentum \cite{SpinT9}, more exotic
quantum states have been unveiled \cite{SpinT8,SpinT9}.

As a convenient experimental knob in atomic and molecular physics, Zeeman
field has been widely used in manipulating spin states \cite{Hspin4,Hspin5},
whereas its impacts on orbital states are usually limited. However, the SO
coupling essentially connects spin and motional degrees of freedom, which
endows orbital states with ability to respond to spin operations and vice
versa. It has been shown that, exploiting SO coupling, target spin states
can be efficiently accessed via relevant manipulations on motional degrees
of freedom \cite{EDSR1,EDSR2,EDSR3,EDSR4}. It is thus anticipated that, in
the presence of SO coupling, the motional character of quantum particles may
be predominantly affected by external Zeeman fields. Given that the
SO-coupled spin-1 quantum gases are naturally subject to both linear and
quadratic Zeeman fields \cite{SpinT1,SpinT2,SpinT3,SpinT4,SpinT5}, an
interesting question is what the respective effects of the two fields on the
atomic orbital and spin dynamics are?

In this paper, we investigate the orbital and spin dynamics of a SO-coupled
spin-1 BEC under the action of both the linear and quadratic Zeeman fields.
The dynamics of the BEC is switched on by a sudden shift of the trapping
potential. It turns out that the Zeeman fields impose crucial impacts on
both the spin and motional degrees of freedom of the BEC. Specifically, the
time-averaged center-of-mass (COM) oscillation and spin polarizations
exhibit remarkable resonant peaks at some special Zeeman field strengths.
The physics underlying this resonant effect can be traced back to the
out-of-phase interference of the dynamical phases carried by different
spin-orbit states. By analyzing the single particle spectrum, the resonant
condition is found to be the level avoided crossing points, which is
summarized as a simple algebraic relation. This relation connects the
strengths of the linear and quadratic Zeeman fields and provides a promising
scheme to calibrate parameters associated with these fields.

\section{System and Hamiltonian}

\label{sec:System and Hamiltonian}

\begin{figure}[tp]
\includegraphics[width=8cm]{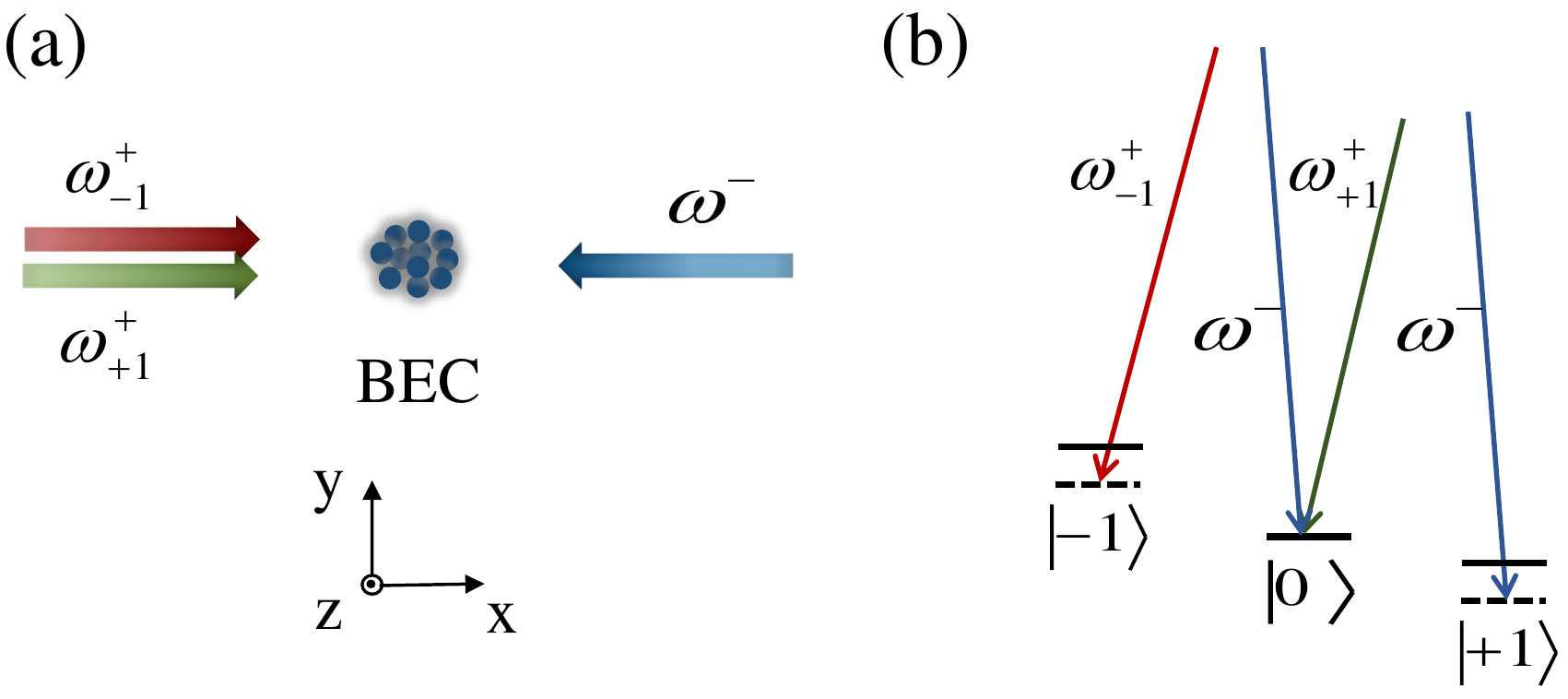}
\caption{(a) Schematic of the system. A spin-1 BEC is illuminated by three
Raman lasers so as to generate SO coupling. (b) The atomic level structure. }
\label{SetupI}
\end{figure}

As illustrated in Fig.~\ref{SetupI}, the system in consideration is similar
as that of Ref. \cite{SpinE1}, where the hyperfine ground states $\left\vert
+1\right\rangle =\left\vert F=1,m_{F}=1\right\rangle $, $\left\vert
0\right\rangle =\left\vert F=1,m_{F}=0\right\rangle $ and $\left\vert
-1\right\rangle =\left\vert F=1,m_{F}=-1\right\rangle $ of $^{87}$Rb atoms
define the three different spin components of the BEC. A magnetic field
along $z$-axis splits the hyperfine sublevels by an energy shift of $\hbar
\omega _{Z}$.\ The pair of counterpropagating laser beams with frequencies $%
\omega ^{-}$ and $\omega _{+1}^{+}$ ($\omega _{-1}^{+}$) induces a
two-photon Raman transition between $\left\vert 0\right\rangle $ and $%
\left\vert +1\right\rangle $ ($\left\vert -1\right\rangle $), and transfers $%
2\hbar k_{r}$ recoil momentum to the atoms at the same time. In the
pseudo-spin-1 basis $\Psi =(\psi _{+1},\psi _{0},\psi _{-1})$, the
single-particle Hamiltonian is written as \cite{SpinT1,SpinT2,SpinT3,SpinT4}%
\begin{equation}
\hat{H}_{S}=\frac{\mathbf{P}^{2}}{2m}+V_{\text{T}}(\mathbf{r})+\mathbf{%
\tilde{\Omega}}(x)\cdot \mathbf{F}+\hbar \delta F_{z}+\hbar \epsilon
F_{z}^{2},  \label{HS1}
\end{equation}%
where $\mathbf{P}^{2}/2m$ and $V_{\text{T}}(\mathbf{r})$ are respectively
the kinetic energy and harmonic trapping potential, $\mathbf{\tilde{\Omega}}%
(x)=\Omega _{R}[\cos (2k_{r}x)\mathbf{e}_{x}-\sin (2k_{r}x)\mathbf{e}_{y}]$
is a space-dependent effective field, $\Omega _{R}$ is the Raman Rabi
frequency, $\mathbf{F}=(F_{x},F_{y},F_{z})$ denotes the spin-1 Pauli
matrices, and $\delta =\omega ^{-}-(\omega _{+1}^{+}+\omega
_{-1}^{+})/2-\omega _{Z}$ contributes the linear Zeeman shift. Note that
besides $\delta $, a quadratic Zeeman field which is not associated with any
spatial direction, $\epsilon =\Delta _{0}+(\omega _{-1}^{+}-\omega
_{+1}^{+})/2$ with $\Delta _{0}$ being the energy shift of state $\left\vert
0\right\rangle $, emerges. Both the two Zeeman terms $\delta $ and $\epsilon
$ can be independently tuned from the positive to the negative by, for
example, varying the frequencies of Raman lasers or the technique of
microwave dressing \cite{Mdressing1,Mdressing2}. Since $\mathbf{\tilde{\Omega%
}}(x)$ plays a role\ only along the spatial $x$ direction, the motional
degrees of freedom along other directions are thus irrelevant as long as we
focus on the physics along the $x$ axis. After integration over $y$ and $z$
degrees of freedom and the unitary transformation $\psi _{\pm
}\longrightarrow \psi _{\pm }e^{\pm 2ik_{r}x}$, we obtain the Hamiltonian in
a form explicitly exhibiting SO coupling,%
\begin{eqnarray}
H_{S} &=&\frac{p_{x}^{2}}{2m}+V(x)+\hbar \Omega F_{x}+\alpha
p_{x}F_{z}+\hbar \delta F_{z}  \notag \\
&&+(\hbar \epsilon +\frac{1}{2}m\alpha ^{2})F_{z}^{2},  \label{HS2}
\end{eqnarray}%
where $V(x)=m\omega ^{2}x^{2}/2$ is the harmonic trapping potential with $%
\omega $ being the trapping frequency in the $x$ direction, $\Omega =\sqrt{2}%
\Omega _{R}/2$ is the transverse-Zeeman potential, and $\alpha =4\hbar
k_{r}/(2m)$ quantifies the SO coupling strength.

Incorporating the interatomic collisional interactions, the dynamics of the
BEC are governed by the Gross-Pitaevskii (G-P) equation%
\begin{equation}
i\hbar \frac{\partial \Psi }{\partial t}=\left( H_{S}+H_{I}\right) \Psi ,
\label{GP}
\end{equation}%
where $H_{I}$ is the mean-field Hamiltonian accounting for the nonlinear
interaction between atoms,%
\begin{equation}
H_{I}=\left(
\begin{array}{ccc}
\Gamma _{+1} & 0 & \Gamma _{-+} \\
0 & \Gamma _{0} & 0 \\
\Gamma _{-+}^{\ast } & 0 & \Gamma _{-1}%
\end{array}%
\right) .
\end{equation}%
Here $\Gamma _{\pm 1}=(c_{0}+c_{2})\left\vert \psi _{\pm 1}\right\vert ^{2}$%
, $\Gamma _{0}=(c_{0}+c_{2}/2)\left\vert \psi _{0}\right\vert ^{2}$, and $%
\Gamma _{-+}=-2c_{2}\psi _{-1}^{\ast }\psi _{+1}$. The coefficients $c_{0}$
and $c_{2}$ describe density-density and spin-spin interaction strengths,
respectively. Note that $c_{0,2}$ can be feasibly tuned through Feshbach
resonances. In the following numerical calculations, we fix $c_{0}\sqrt{%
m/\hbar ^{3}\omega }=0.05$ and take the typical ratio $c_{2}/c_{0}=-0.005$
for $^{87}$Rb.

\section{Single particle spectrum}

\label{sec:Single particle spectrum}

\begin{figure}[tp]
\includegraphics[width=8cm]{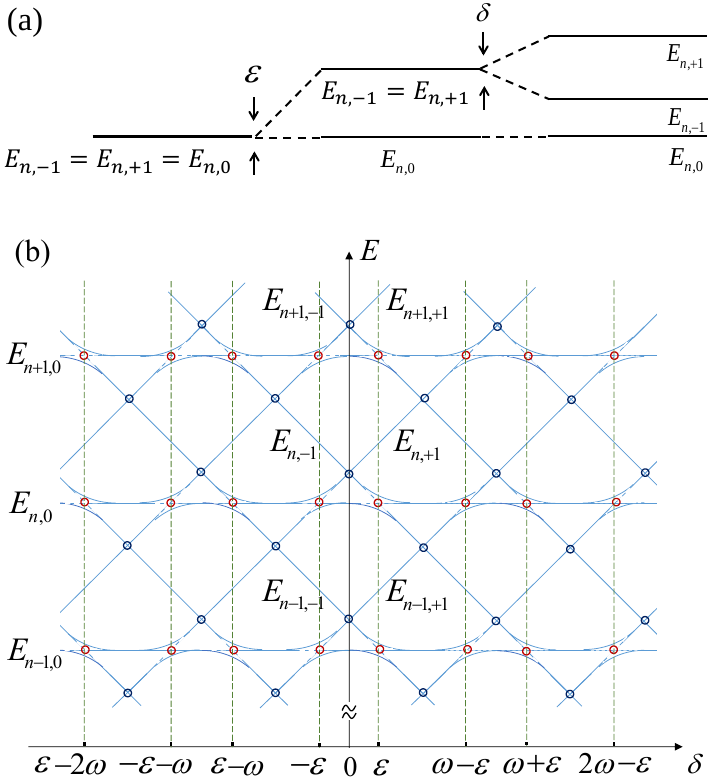}
\caption{(a) Schematic energy levels of the $n$th orbit and their splitting
under the action of the quadratic and linear Zeeman fields. (b) Schematic
illustration of the single-particle spectrum (blue solid curve) as functions
of the linear Zeeman field $\protect\delta $ with certain fixed $\protect%
\epsilon $. The red and blue circles indicate the level avoided crossing and
level crossing points, respectively. The green dashed lines are
intentionally added to pinpoint the resonant values of $\protect\delta $.}
\label{Levelcross}
\end{figure}

We first analyze the single-particle spectrum of the system. Notice that in
the absence of the transverse potential ($\Omega =0$), the Hamiltonian~(\ref%
{HS2}) is exactly solvable, giving rise to the eigenstates%
\begin{equation}
\left\vert \psi _{n,\chi }\right\rangle =\left\vert \psi _{n}^{\chi
}\right\rangle \left\vert \chi \right\rangle ,  \label{PST}
\end{equation}%
where the spin part $\left\vert \chi \right\rangle $ is the eigenstate of
the spin operator $F_{z}$, obeying $F_{z}\left\vert \chi \right\rangle =\chi
\left\vert \chi \right\rangle $ with $\chi =0,\pm 1$, and the orbital part
satisfies $\left\vert \psi _{n}^{\chi }\right\rangle \equiv \exp (-i\chi
m\alpha x/\hbar )\left\vert \phi _{n}\right\rangle $. Here $\left\vert \phi
_{n}\right\rangle $ is the $n$th eigenstate of a harmonic oscillator whose
oscillation frequency is $\omega $. The eigenvalues of states (\ref{PST})
are given by
\begin{equation}
E_{n,\chi }=n\hbar \omega -\chi (\hbar \delta -\chi \hbar \epsilon ).
\label{EN1}
\end{equation}%
It is thus clear that, due to the spin-1 nature of the BEC, the spectra are
grouped as three different branches labeled by $\chi $, each of which
contains a series of equally-distributed orbital levels. Moreover, without
the linear and quadratic Zeeman fields, the orbital levels with the same
quantum number $n$ appear to be triply degenerate with respect to the spin
variation $\chi \longrightarrow \chi \pm 1$. A nonzero quadratic Zeeman
field opens an energy splitting between $E_{n,0}$ and $E_{n,\pm 1}$ by $%
\hbar \epsilon $, and the degeneracy of the doublet $E_{n,+1}$ and $E_{n,-1}$
is lifted by the linear Zeeman field [see Fig.~\ref{Levelcross}(a) for
illustration]. It is to be noted that, by further increasing the quadratic
and linear Zeeman fields, there exists a possibility that the eigenstates
with different spin and orbital quantum numbers become degenerate again,
namely $E_{n,\chi }=E_{n+k,\chi ^{\prime }}$, where $k$ is a nonzero integer
and $\chi \neq \chi ^{\prime }$. This degeneracy may dramatically affect the
dynamics of the BEC, as will be clarified in Sec.~\ref{sec:Dynamical
resonance}.

Let us now turn on the transverse potential ($\Omega \neq 0$) and inspect
its influences on the spectrum. Since the transverse term $\hbar \Omega
F_{x} $ does not commute with the Hamiltonian, no exact solution exists. We
work on the regime where $\Omega /\omega \ll 1$ so that the transverse
potential can be treated perturbatively. Based on the standard perturbation
theory, the eigenstates, which are accurate up to first order in $\Omega $,
are summarized as
\begin{eqnarray}
\left\vert \tilde{\psi}_{n,\chi }\right\rangle &=&\sum_{n^{\prime
}=1}^{\infty }\left( C_{\chi ,-1}^{n,n^{\prime }}\left\vert \psi _{n^{\prime
}}^{-1}\right\rangle \left\vert -1\right\rangle +C_{\chi ,0}^{n,n^{\prime
}}\left\vert \psi _{n^{\prime }}^{0}\right\rangle \left\vert 0\right\rangle
\right.  \notag \\
&&\left. +C_{\chi ,+1}^{n,n^{\prime }}\left\vert \psi _{n^{\prime
}}^{+1}\right\rangle \left\vert +1\right\rangle \right) ,  \label{DPST}
\end{eqnarray}%
where the detailed expressions of $C_{\chi ,\chi ^{\prime }}^{n,n^{\prime }}$
are given in the Appendix. It can be seen that the states $\left\vert \tilde{%
\psi}_{n,\chi }\right\rangle $ are no longer spin-orbit separable but in a
form that spin and orbital parts are dressed together. Without the
transverse potential ($\Omega =0$), we have $C_{\chi ,\chi ^{\prime
}}^{n,n^{\prime }}=\delta _{\chi ,\chi ^{\prime }}\cdot \delta _{n,n^{\prime
}}$ and the dressed state reduces to the bare one, $\left\vert \tilde{\psi}%
_{n,\chi }\right\rangle =\left\vert \psi _{n,\chi }\right\rangle $.
Observing $\left\langle \psi _{n,\chi }\right\vert F_{x}\left\vert \psi
_{n^{\prime },\chi ^{\prime }}\right\rangle =\sqrt{2}(\delta _{n,n^{\prime
}}\delta _{\chi ,\chi ^{\prime }+1}+\delta _{n,n^{\prime }}\delta _{\chi
,\chi ^{\prime }-1})/2$, the first order corrections of eigenvalues are
generally zero so that corresponding eigenenergies remain the same as those
without transverse potential, $\tilde{E}_{n,\chi }=E_{n,\chi }$. However,
when energy levels differing by one unit of spin angular momentum get close
to each other, saying $\left\vert E_{n,0}-E_{n+k,\pm 1}\right\vert \ll $ $%
\hbar \omega $, the coupling between them is intensively enhanced, leading
to the break down of the non-degenerate perturbation formula. Employing a
degenerate perturbation method, an avoided crossing between energy levels
with $\left\vert \chi -\chi ^{\prime }\right\vert =1$ appears, producing
\begin{equation}
\left\{
\begin{array}{c}
\tilde{E}_{n+k,0}=(n+k)\hbar \omega +\frac{\sqrt{2}}{2}\hbar \Omega
\left\vert \eta \right\vert \\
\tilde{E}_{n,+1}=(n+k)\hbar \omega -\frac{\sqrt{2}}{2}\hbar \Omega
\left\vert \eta \right\vert \\
\;\tilde{E}_{n,-1}=E_{n,-1}%
\end{array}%
\right. ,  \label{EIN1}
\end{equation}%
for $E_{n,+1}=E_{n+k,0}$, and
\begin{equation}
\left\{
\begin{array}{c}
\tilde{E}_{n,+1}=E_{n,+1} \\
\tilde{E}_{n+k,0}=(n+k)\hbar \omega +\frac{\sqrt{2}}{2}\hbar \Omega
\left\vert \eta \right\vert \\
\tilde{E}_{n,-1}=(n+k)\hbar \omega -\frac{\sqrt{2}}{2}\hbar \Omega
\left\vert \eta \right\vert%
\end{array}%
\right. ,  \label{EIN2}
\end{equation}%
for $E_{n,-1}=E_{n+k,0}$, with transverse-potential-induced splitting $\sqrt{%
2}\hbar \Omega \left\vert \eta \right\vert $ and $\eta =\left\langle \psi
_{n}^{+1}\right\vert \left. \psi _{n+k}^{0}\right\rangle $ [see Fig.~\ref%
{Levelcross}(b) for illustration]. Note that since $F_{x}$ does not couple
states with spin angular momentum $+1$ and $-1$, the non-degenerate
perturbation theory still applies\ for the case of $E_{n,+1}=E_{n+k,-1}$, at
which a level crossing occurs instead.\ \ \

\section{Dynamical resonance}

\label{sec:Dynamical resonance}

Armed with the knowledge of eigenstates and eigenenergies, we are in the
right stage to study the collective dynamics. We first focus on the COM
motion of the condensate subject to a sudden shift of the harmonic trapping
potential. It is well known that for a regular BEC without SO coupling, the
COM motion turns out to be a sinusoidal oscillation whose period depends
only on the trapping frequency and is not affected by other parameters such
as nonlinearity, shifting distance, and external Zeeman fields \cite{HCOM}.
The SO coupling, on the other hand, embeds the spin character of the BEC
into its motional degrees of freedom \cite{ASOD2,ASOD3,ASOD4}. In view of
this, the COM motion here is expected to respond to typical spin
manipulations, which for instance, can be achieved by applying effective
magnetic fields such as the linear and quadratic Zeeman fields.
\begin{figure}[tp]
\includegraphics[width=8cm]{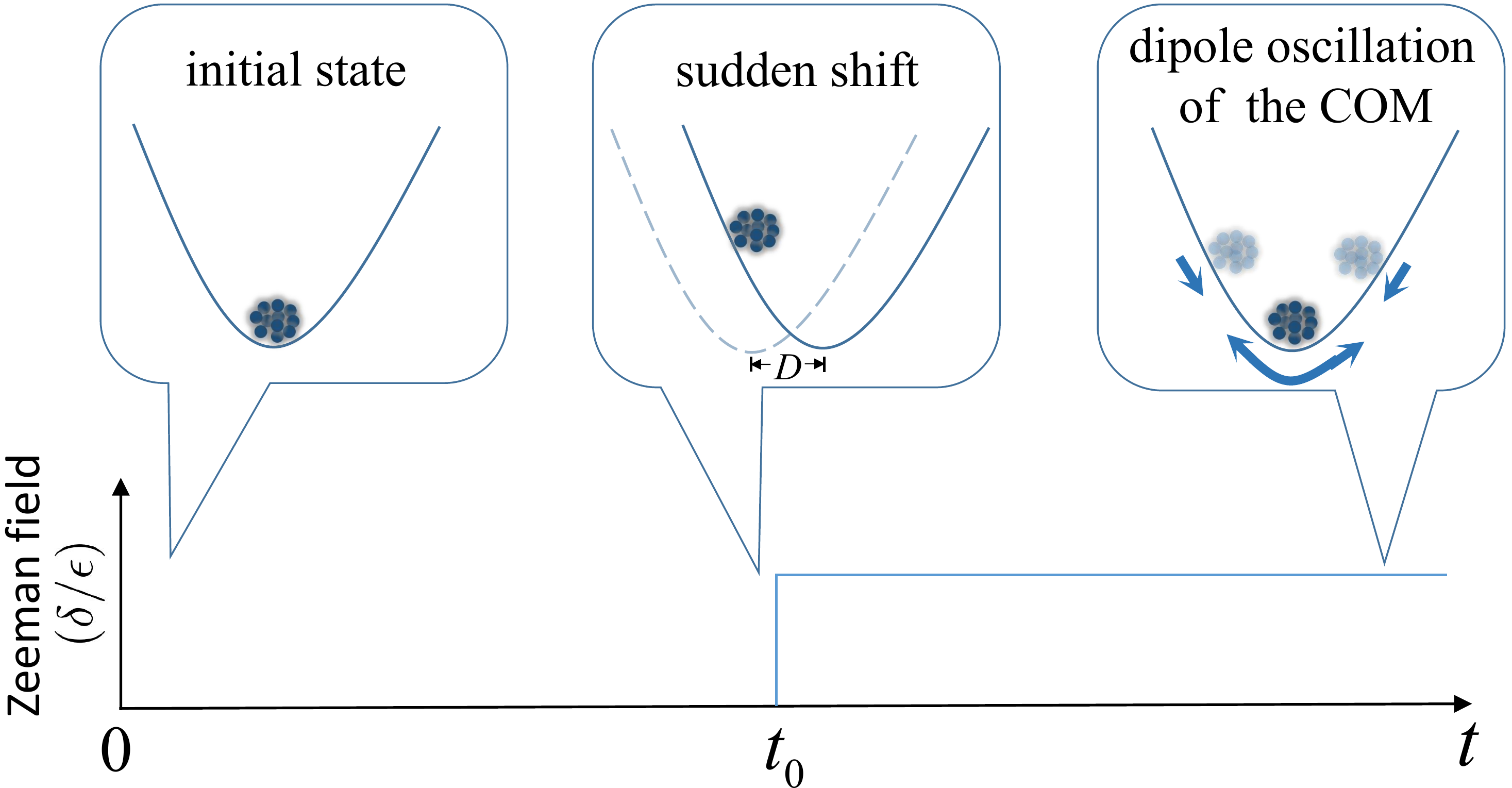}
\caption{Schematic description of the proposed dynamical scheme for the
Zeeman resonance. Top panel: Trapping potential and the corresponding COM
motion of the BEC at different times. Bottom panel: Timing of the external
Zeeman fields.}
\label{SetupII}
\end{figure}

We assume the external Zeeman fields are switched off initially ($\delta
=\epsilon =0$), and the BEC is prepared in a given state of the lowest
orbital level, say $\left\vert \Psi (0)\right\rangle =\sum\nolimits_{\chi
=-1}^{+1}C_{\chi }\left\vert \psi _{0}^{\chi }\right\rangle \left\vert \chi
\right\rangle $, with $C_{\chi }$ being some superposition coefficient. The
dynamics is activated at some time $t_{0}$ by a sudden shift of the trapping
potential \cite{ASOD2,ASOD4}. Moreover, the trap shift is accompanied by an
abruptly applied linear and quadratic Zeeman fields, as schematically
illustrated in Fig.~\ref{SetupII}. Given this, the wavefunction for $t>t_{0}$
can be expanded in terms of the eigenstates $\left\vert \tilde{\psi}_{n,\chi
}\right\rangle $ as
\begin{equation}
\left\vert \Psi (t)\right\rangle =\sum_{n=1}^{\infty }\sum_{\chi
=-1}^{1}A_{n,\chi }e^{\frac{i}{\hbar }p_{x}D}\left\vert \tilde{\psi}_{n,\chi
}\right\rangle e^{-\frac{i}{\hbar }\tilde{E}_{n,\chi }},  \label{WF}
\end{equation}%
where $A_{n,\chi }=\left\langle \tilde{\psi}_{n,\chi }\right\vert \exp
(-ip_{x}D/\hbar )\left\vert \Psi (0)\right\rangle $ with $D$ being the
shifting distance. With this wave function, the time evolution of the COM is
expressed as%
\begin{eqnarray}
\left\langle x(t)\right\rangle &\!\!\!=\!\!\!&\sum_{n=1}^{\infty
}\sum_{n^{\prime }=1}^{\infty }\sum_{\chi =-1}^{1}\sum_{\chi ^{\prime
}=-1}^{1}\sqrt{\frac{m\omega }{\hbar }}R_{\chi ,\chi ^{\prime
}}^{n,n^{\prime }}e^{-\frac{i}{\hbar }\left( \tilde{E}_{n,\chi }-\tilde{E}%
_{n^{\prime },\chi ^{\prime }}\right) t}  \notag \\
&&+D,  \label{X1}
\end{eqnarray}%
where $\left\langle \cdot \cdot \cdot \right\rangle $ stands for the spacial
average over condensate wave function, and we have introduced the
dimensionless term $R_{\chi ,\chi ^{\prime }}^{n,n^{\prime }}=\sqrt{\hbar
/m\omega }A_{n^{\prime },\chi ^{\prime }}^{\ast }A_{n,\chi }\left\langle
\tilde{\psi}_{n^{\prime },\chi ^{\prime }}\right\vert x\left\vert \tilde{\psi%
}_{n,\chi }\right\rangle $. Note that in Eq.~(\ref{X1}), while terms
multiplied by the dynamical phase factors, $\exp [-i(\tilde{E}_{n,\chi }-%
\tilde{E}_{n^{\prime },\chi ^{\prime }})t/\hbar ]$, are responsible for the
time-dependent oscillation, the shifting distance $D$ appearing in the last
term represents a constant equilibrium position around which the BEC
oscillates. As we are only interested in the dynamical part of the
oscillation, it is more convenient to focus on the redefined COM motion in
which the constant shifting distance is deducted, i.e., $\left\langle
\overline{x(t)}\right\rangle =\left\langle x(t)\right\rangle -D$. Inspired
by the fact that quantum particles in a given state are essentially nonlocal
in their spacial dimensions, it is expected that there may be some physical
information nonlocally hidden in the time dimension of wave functions. This
motivate us to investigate the time-averaged quantity
\begin{equation}
Q(\mathcal{O})=\frac{1}{T}\left\vert \int_{t_{0}}^{t_{0}+T}\left\langle
\mathcal{O}\right\rangle dt\right\vert ,  \label{TE}
\end{equation}%
where $T$ is a long-time span and $\mathcal{O}$ is the physical observable
over which the average is taken \cite{ASOD5}. Equation (\ref{TE}) can be
treated as a kind of coarse-grained averaging, as the dynamical details at
any specific time become irrelevant. Nevertheless, potential dynamical
effects accumulated through a long-time evolution are remarkably highlighted
under this framework.

\begin{figure}[tp]
\includegraphics[width=8.5cm]{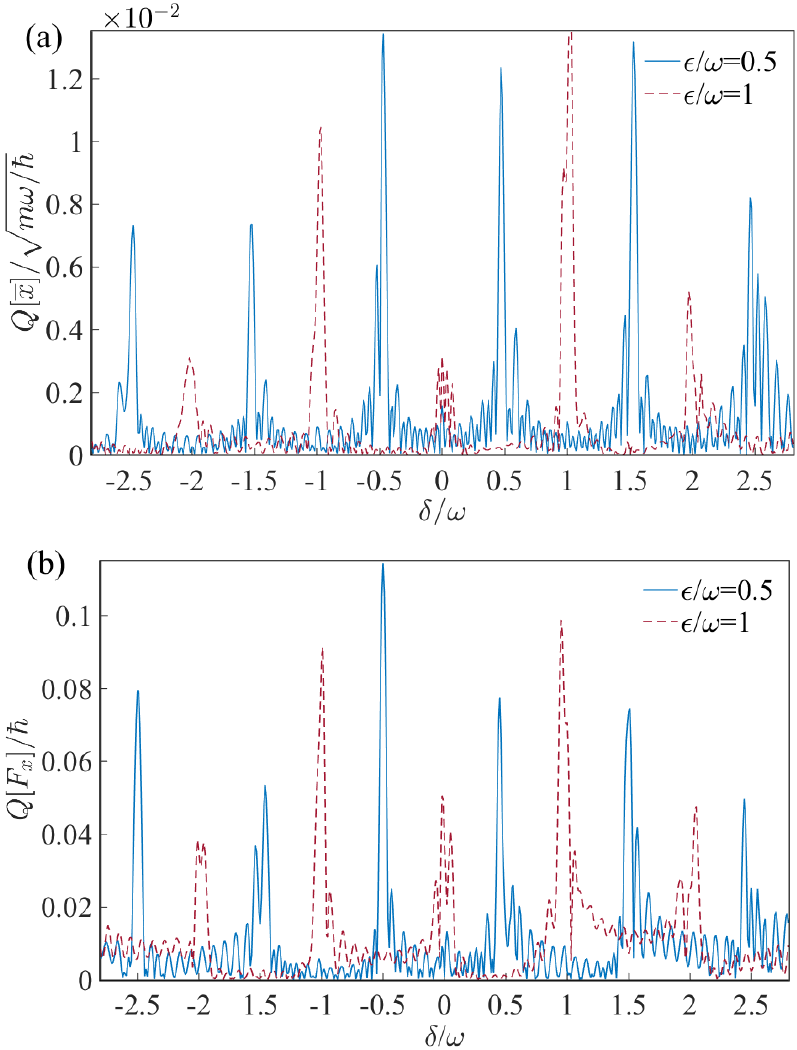}
\caption{The long-time-averaged quantity (a) $Q[\overline{x(t)}]$ and (b) $%
Q[F_{x}(t)]$ as functions of the linear Zeeman field $\protect\delta $ for $%
\protect\epsilon /\protect\omega =0.5$ (blue-solid curve) and $\protect%
\epsilon /\protect\omega =1$ (red-dashed curve), with the initial state $%
\left\vert \Psi (0)\right\rangle =(\left\vert \protect\psi %
_{0}^{-1}\right\rangle \left\vert -1\right\rangle +\left\vert \protect\psi %
_{0}^{0}\right\rangle \left\vert 0\right\rangle +\left\vert \protect\psi %
_{0}^{+1}\right\rangle \left\vert +1\right\rangle )/\protect\sqrt{3}$. The
other parameters are $T=70/\protect\omega $, $D/\protect\sqrt{m\protect%
\omega /\hbar }=2$, $\Omega /\protect\omega =0.05$ and $\protect\alpha =2$.}
\label{Peak1}
\end{figure}

As depicted in Fig.~\ref{Peak1}(a), the long-time-averaged COM motion, $Q[%
\overline{x(t)}]$, as a function of $\delta $ for various $\epsilon $ with
the initial state $\left\vert \Psi (0)\right\rangle =(\left\vert \psi
_{0}^{-1}\right\rangle \left\vert -1\right\rangle +\left\vert \psi
_{0}^{0}\right\rangle \left\vert 0\right\rangle +\left\vert \psi
_{0}^{+1}\right\rangle \left\vert +1\right\rangle )/\sqrt{3}$, is obtained
by numerically solving the G-P equation (\ref{GP}). An intriguing finding is
that a series of resonant peaks are formed at some specific linear Zeeman
fields, whose values appear to be affected by the strength of the quadratic
Zeeman field. That is, for $\epsilon /\omega =1$, the resonance occurs at
integer values of $\delta /\omega $, whereas for $\epsilon /\omega =0.5$,
the resonant points of $\delta /\omega $ become half-integer. This phenomena
can be viewed as a consequence of the out-of-phase interference among
different spin-orbit states.

To understand this clearly, attention should be turned to the dynamical
phase factors $\exp [-i(\tilde{E}_{n,\chi }-\tilde{E}_{n^{\prime },\chi
^{\prime }})t/\hbar ]$ and the corresponding terms $R_{\chi ,\chi ^{\prime
}}^{n,n^{\prime }}$ in Eq.~(\ref{X1}). Note that the dynamical phases in the
form of $(\tilde{E}_{n,\chi }-\tilde{E}_{n^{\prime },\chi ^{\prime
}})t/\hbar $ with $\chi \neq \chi ^{\prime }$ strongly\ depend on both the
two Zeeman fields $\delta $ and $\epsilon $, whereas those possessing a
single spin subscript $\chi $, namely $-i(\tilde{E}_{n,\chi }-\tilde{E}%
_{n^{\prime },\chi })t/\hbar $, do not. It is thus expected that the former
should play the key role in any Zeeman-field-related responses. In fact, for
general values of $\delta $ and $\epsilon $, the diagonal terms $R_{\chi
,\chi }^{n,n}$ are negligibly small since they are shown to be proportional
to $(\Omega /\omega )^{2}$ (See Appendix for details), and it is the energy
differences, $\tilde{E}_{n,\chi }-\tilde{E}_{n^{\prime },\chi ^{\prime }}$ ($%
n\neq n^{\prime }$ and $\chi \neq \chi ^{\prime }$), which are on the order
of a few $\hbar \omega $, that dominates the time evolution of the BEC. As a
result, the dynamical parts in Eq.~(\ref{X1}) oscillate fast over time,
making $Q[\overline{x(t)}]$ tend to vanish due to the out-of-phase
interference. However, tuning Zeeman fields to the level avoided crossing
point with $E_{n,0}=E_{n+k,\pm 1}$, we get a maximally minimized energy
difference, satisfying $\left\vert \tilde{E}_{n,\chi }-\tilde{E}_{n^{\prime
},\chi ^{\prime }}\right\vert /\hbar \omega =\sqrt{2}\left\vert \eta
\right\vert \Omega /\omega \ll 1$, which dramatically slows down the time
oscillation of the phase factors. Hence, the out-of-phase interference is
suppressed to the largest extent, giving rise to a considerable non-zero
contribution to $Q[\overline{x(t)}]$. It follows that the level avoided
crossing point, at which the Zeeman fields $\delta $ and $\epsilon $ satisfy
\begin{equation}
\delta \pm \epsilon =k\omega ,  \label{RC}
\end{equation}%
is nothing but the point at which the dynamical resonance occurs. The
resonant condition Eq.~(\ref{RC}) is the main result of this paper.

\begin{figure}[tp]
\includegraphics[width=8cm]{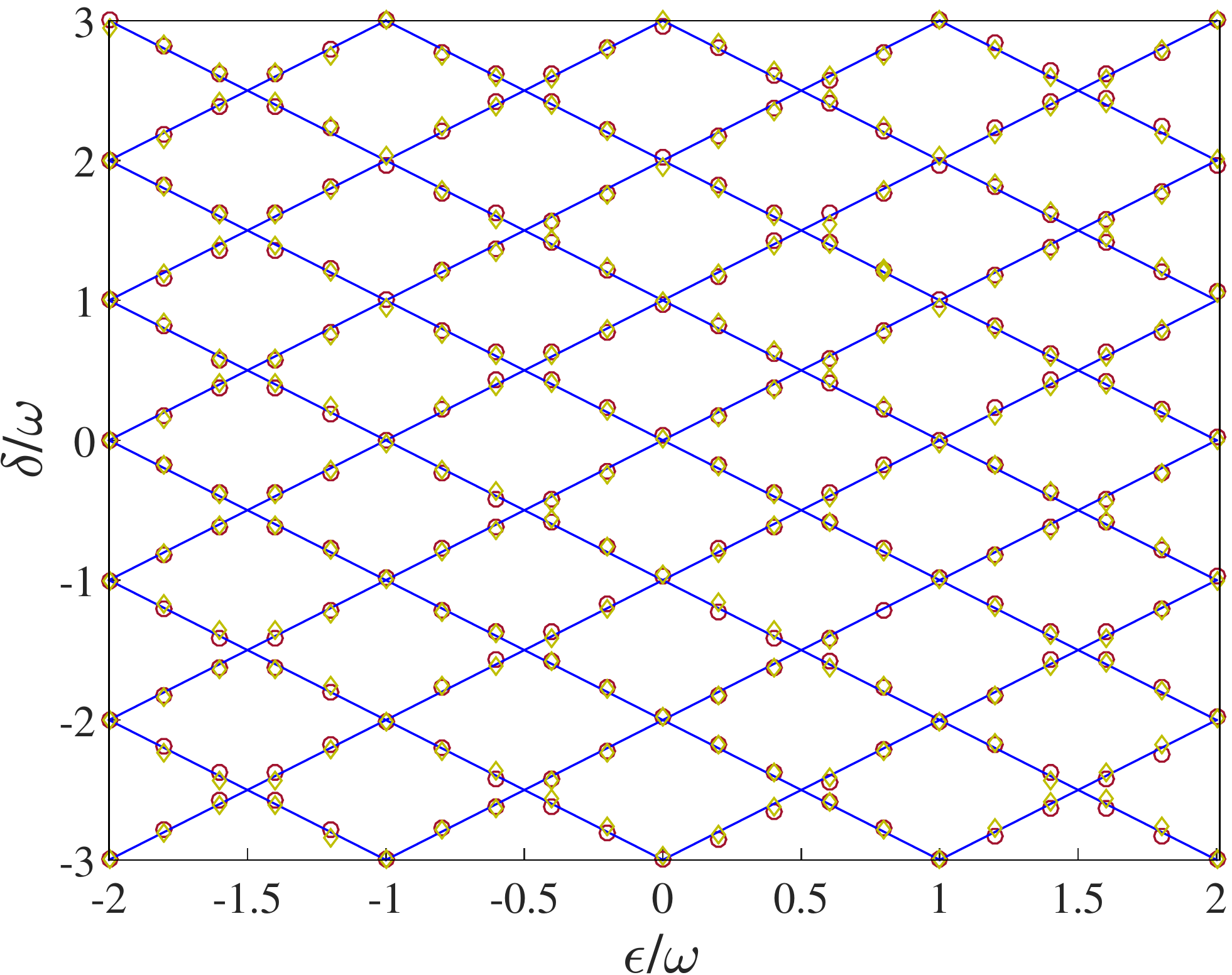}
\caption{Plot of resonance condition in the $\protect\epsilon -\protect%
\delta $ plane. The blue solid curves correspond to the analytical relation
in Eq.~(\protect\ref{RC}). The red circles (yellow diamonds) come from
numerical results of $Q[\overline{x(t)}]$ ($Q[F_{x}(t)]$) obtained by
solving the G-P equation (\protect\ref{GP}). The initial state and other
parameters used in the numerical calculations are the same as those in Fig.~%
\protect\ref{Peak1}. }
\label{resonance}
\end{figure}

Along this reasoning, it seems that there should exist similar resonant
peaks at the level crossing point with $E_{n,+1}=E_{n+k,-1}$ as well [See,
for example, the blue circles in Fig.~\ref{Levelcross}(b)]. However, a
straightforward calculation shows that, at this point, $R_{+1,-1}^{n,n+k}$
is on the order of $(\Omega /\omega )^{2}A_{n^{\prime },\chi ^{\prime
}}^{\ast }A_{n,\chi }$, which turns out to be vanishingly small in the
perturbation regime. Thus, the suppression of out-of-phase interference can
generate little contribution to $Q[\overline{x(t)}]$, resulting in the
absence of the expected dynamical resonance. Figure \ref{resonance} plots
the coordinates of resonant peaks in the $\epsilon -$ $\delta $ plane, which
is obtained by numerically solving the G-P equation (\ref{GP}). It is shown
that the numerical results are in quantitative agreement with Eq.~(\ref{RC}%
). The resonance condition in Eq.~(\ref{RC}) is simple and quite generic in
the sense that it bridges between the linear and quadratic Zeeman fields via
only the trapping frequency $\omega $, and is independent of other
parameters such as the shifting distance $D$, the time span $T$, and the SO
coupling strength $\alpha $. This property offers interesting opportunities
for the Zeeman-fields-based quantum metrology.

Instead of responding linearly to the Zeeman fields, as is known for systems
without SO coupling, the spin polarization here may exhibit similar resonant
behavior. The physics follows that of the COM motion. Invoking the wave
function in Eq.~(\ref{WF}), the $i$th\ component ($i=x,y,z$) of the spin
polarization is written as%
\begin{equation}
\left\langle F_{i}(t)\right\rangle =\sum_{n=1}^{\infty }\sum_{n^{\prime
}=1}^{\infty }\sum_{\chi =-1}^{1}\sum_{\chi ^{\prime }=-1}^{1}\hbar R_{\chi
,\chi ^{\prime }}^{(i),n,n^{\prime }}e^{-\frac{i}{\hbar }\left( \tilde{E}%
_{n,\chi }-\tilde{E}_{n^{\prime },\chi ^{\prime }}\right) t},
\end{equation}%
with $R_{\chi ,\chi ^{\prime }}^{(i),n,n^{\prime }}=A_{n^{\prime },\chi
^{\prime }}^{\ast }A_{n,\chi }\left\langle \tilde{\psi}_{n^{\prime },\chi
^{\prime }}\right\vert F_{i}\left\vert \tilde{\psi}_{n,\chi }\right\rangle
/\hbar $. As described above, one of the key points of the Zeeman resonance
lies in the fact that the off-diagonal terms $R_{\chi ,\chi ^{\prime
}}^{(i),n,n^{\prime }}$ dominate over the diagonal ones $R_{\chi ,\chi
}^{(i),n,n\text{ }}$outside the level avoided crossing points. This
motivates us to focus on the spin polarization along the transverse
directions (i.e., directions in the $x-y$ plane), since in this case $%
R_{\chi ,\chi }^{(i),n,n}$ is negligible compared to $R_{\chi ,\chi ^{\prime
}}^{(i),n,n^{\prime }}$ in the sense that $R_{\chi ,\chi }^{(i),n,n}/R_{\chi
,\chi ^{\prime }}^{(i),n,n^{\prime }}\sim \Omega /\omega $. Following the
same derivation as that used in analyzing the COM motion, we can reproduce
the resonance condition in Eq.~(\ref{RC}) straightforwardly. Figure \ref%
{Peak1}(b) shows the numerical results of $Q[F_{x}(t)]$ as a function of $%
\delta $ for various $\epsilon $, whose peak positions are well described by
Eq.~(\ref{RC}). More numerical results of peak positions in the $\epsilon -$
$\delta $ plane are shown in Fig.~\ref{resonance}, which agree with Eq.~(\ref%
{RC}) as expected.

It is worth noting that, our discussion about the proposed Zeeman resonance
is not affected by different choices of the initial state $\left\vert \Psi
(0)\right\rangle $, provided that it is a superposition of the three Zeeman
sublevels $\left\vert +1\right\rangle $, $\left\vert 0\right\rangle $, and $%
\left\vert -1\right\rangle $. Easy to be satisfied in the current experiment
with cold atoms, this constraint on $\left\vert \Psi (0)\right\rangle $
guarantees the off-diagonal terms $R_{\chi ,\chi ^{\prime }}^{n,n^{\prime }}$
and $R_{\chi ,\chi ^{\prime }}^{(x),n,n^{\prime }}$\ nonzero, which is
necessary to support visible resonant peaks.
\begin{figure}[tp]
\includegraphics[width=8.5cm]{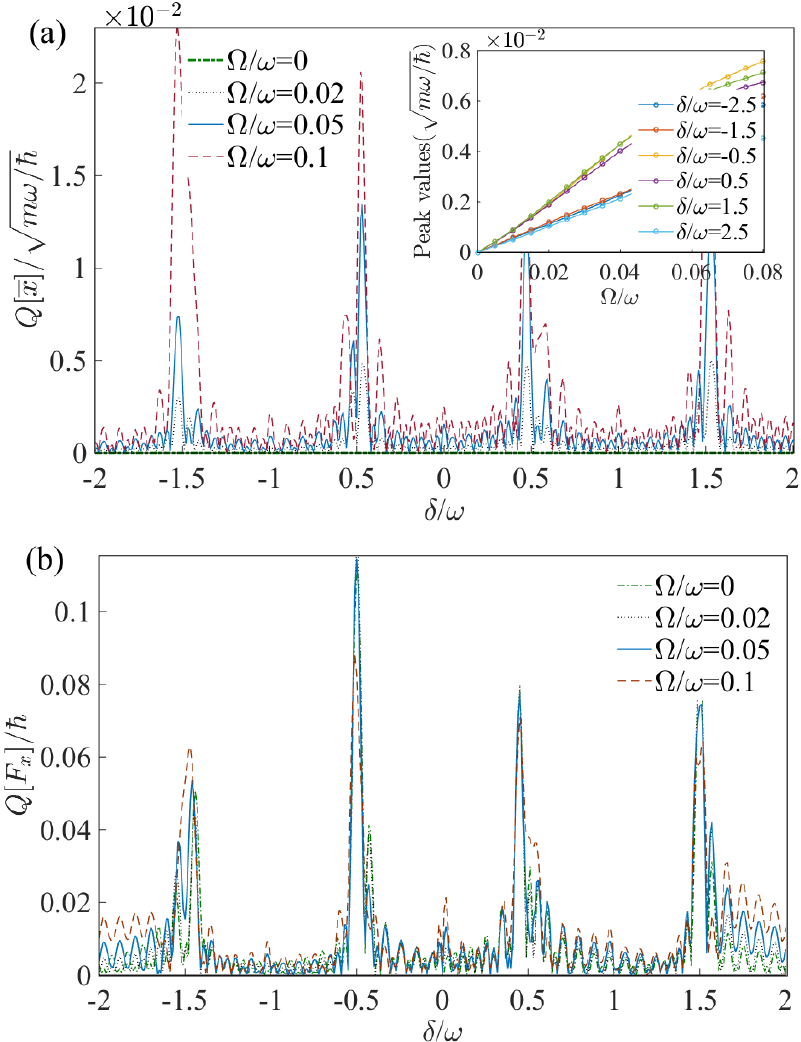}
\caption{The long-time-averaged quantity, $Q[\overline{x(t)}]$ (a) and $%
Q[F_{x}(t)]$ (b), as functions of the linear Zeeman field $\protect\delta $
for $\protect\epsilon /\protect\omega =0.5$ (blue-solid curve) and different
transverse potentials $\Omega $. Inset: The peak values of $Q[\overline{x(t)}%
]$ as a function of $\Omega /\protect\omega $. The initial state and the
other parameters are the same as those in Fig.~\protect\ref{Peak1}. }
\label{Peak2}
\end{figure}

We emphasize that the although the transverse potential $\Omega $ is not
explicitly involved in Eq.~(\ref{RC}), it plays a significant role in
inducing the dynamical Zeeman resonance of the COM motion. It is easy to
check that, in the absence of $\Omega $, the off-diagonal terms, $R_{\chi
,\chi ^{\prime }}^{n,n^{\prime }}$ with $\chi \neq \chi ^{\prime }$, vanish,
owing to the orthogonality between different spin states. This further
erases the corresponding phase factors, $\exp [-i(\tilde{E}_{n,\chi }-\tilde{%
E}_{n^{\prime },\chi ^{\prime }})t/\hbar ]$, in Eq.~(\ref{X1}) so that the
dynamical resonant effect disappears. A nonzero transverse potential, on the
other hand, dresses orbital states in different spin branches, as described
by Eq.~(\ref{DPST}). This renders $R_{\chi ,\chi ^{\prime }}^{n,n^{\prime }}$
acquire finite values and thus validates the resonance condition in Eq.~(\ref%
{RC}). In Fig.~\ref{Peak2}(a), we plot $Q[\overline{x(t)}]$ versus $\delta $
for different $\Omega $ with $\epsilon /\omega =0.5$. This figure shows
that, each peak of $Q[\overline{x(t)}]$ increases in height as $\Omega $
increases. Especially for $\Omega =0$, no peaks can be found. Indeed, around
the level avoided crossing point, the leading order of terms $R_{\chi ,\chi
^{\prime }}^{n,n^{\prime }}$ are shown to be $\Omega /\omega $. This signals
that the peak values of $Q[\overline{x(t)}]$ may scale as $\Omega $ when the
transverse potential is weak enough. In the inset of Fig. \ref{Peak2}(a), we
numerically plot various peak values of $Q[\overline{x(t)}]$ as functions of
$\Omega /\omega $. It is found that these peak values can be well described
by linear functions of $\Omega $ for $\Omega /\omega \lesssim 0.03$.
Interestingly, in contract to the COM motion, the resonant peaks of spin
polarizations appear to have no explicit dependence on $\Omega $, and they
persist even for $\Omega =0$ [see Fig.~\ref{Peak2}(b)]. This is because $%
F_{x}$ couple states with different spin angular momentum, yielding $R_{\chi
,\chi ^{\prime }}^{(x),n,n^{\prime }}\neq 0$, regardless of the explicit
value of $\Omega $.

\section{Conclusions}

\label{sec:Conclusions}

In conclusion, we have investigated the orbital and spin dynamics of a
SO-coupled spin-1 BEC, and unraveled a Zeeman-field-induced resonant effect
in this system. The resonant signature is encoded in the time-averaged COM
oscillation and spin polarizations, which exhibit remarkable peaks when the
Zeeman fields are tuned to certain strengths. The underlying physics behind
this resonance can be attributed to the out-of-phase interference of the
dynamical phases carried by different SO states. We have also derived an
analytical expression for the resonant condition. This expression set a
connection between the linear and quadratic Zeeman fields, and may thus
facilitate applications in quantum information and quantum precision
measurement.

\acknowledgments

This work is supported partly by the National Key R\&D Program of China
under Grant No.~2017YFA0304203; the NSFC under Grants No.~11674200 and
No.~11804204; and Shanxi ``1331 Project" Key Subjects Construction.

\vbox{\vskip1cm} \appendix

\section{perturbation calculations}

In this Appendix, we provide the detailed derivation of the eigenstates in
Eq.~(\ref{DPST}) and eigenenergies Eqs.~(\ref{EIN1}) and (\ref{EIN2}) of the
main text, based on the perturbation theory. For general parameters, the
unperturbed eigenenergies are non-degenerate and thus the non-degenerate
perturbation formula applies. The eigenstates, which are accurate up to
first order in $\hbar \Omega $, are obtained as
\begin{eqnarray}
\left\vert \tilde{\psi}_{n,\pm 1}\right\rangle &=&\left\vert \psi _{n}^{\pm
1}\right\rangle \left\vert \pm 1\right\rangle  \label{NCS1} \\
&&+\frac{\sqrt{2}\hbar \Omega }{2}\sum_{n^{\prime }=1}^{\infty }\frac{%
\left\langle \psi _{n^{\prime }}^{0}\right. \left\vert \psi _{n}^{\pm
1}\right\rangle }{(n-n^{\prime })\hbar \omega \mp \hbar \delta +\hbar
\varepsilon }\left\vert \psi _{n^{\prime }}^{0}\right\rangle \left\vert
0\right\rangle ,  \notag \\
\left\vert \tilde{\psi}_{n,0}\right\rangle &=&\left\vert \psi
_{n}^{0}\right\rangle \left\vert 0\right\rangle  \label{NCS2} \\
&&+\frac{\sqrt{2}\hbar \Omega }{2}\sum_{n^{\prime }=1}^{\infty }\frac{%
\left\langle \psi _{n^{\prime }}^{+1}\right. \left\vert \psi
_{n}^{0}\right\rangle }{(n-n^{\prime })\hbar \omega +\hbar \delta -\hbar
\varepsilon }\left\vert \psi _{n^{\prime }}^{+1}\right\rangle \left\vert
+1\right\rangle  \notag \\
&&+\frac{\sqrt{2}\hbar \Omega }{2}\sum_{n^{\prime }=1}^{\infty }\frac{%
\left\langle \psi _{n^{\prime }}^{-1}\right. \left\vert \psi
_{n}^{0}\right\rangle }{(n-n^{\prime })\hbar \omega -\hbar \delta -\hbar
\varepsilon }\left\vert \psi _{n^{\prime }}^{-1}\right\rangle \left\vert
-1\right\rangle .  \notag
\end{eqnarray}%
The corresponding eigenenergies are $\tilde{E}_{n,\chi }=E_{n,\chi }+\hbar
\Omega \left\langle \psi _{n,\chi }\right\vert F_{x}\left\vert \psi _{n,\chi
}\right\rangle =E_{n,\chi }$. With the states in Eqs.~(\ref{NCS1}) and (\ref%
{NCS2}), we can readily make the following estimation of orders: $R_{\chi
,\chi }^{n,n}/(A_{n^{\prime },\chi ^{\prime }}^{\ast }A_{n,\chi })\sim
(\Omega /\omega )^{2}$, $R_{0,\pm 1}^{n,n^{\prime }}/(A_{n^{\prime },\chi
^{\prime }}^{\ast }A_{n,\chi })\sim \Omega /\omega $, $R_{\chi ,\chi
}^{(x/y),n,n}/(A_{n^{\prime },\chi ^{\prime }}^{\ast }A_{n,\chi })\sim
\Omega /\omega $, and $R_{0,\pm 1}^{(x/y),n,n^{\prime }}/(A_{n^{\prime
},\chi ^{\prime }}^{\ast }A_{n,\chi })\sim 1$.

However, when the energy levels are tuned to the level avoided crossing
point where $\left\vert E_{n,\chi }-E_{n+k,\chi ^{\prime }}\right\vert \ll $
$\hbar \omega $, we should employ the degenerate perturbation theory.
Assuming, for instance, $E_{n,+1}=E_{n+k,0}$, the degenerate subspace is
spanned by $\left\vert \psi _{n,+1}\right\rangle $ and $\left\vert \psi
_{n+k,0}\right\rangle $. The secular equation of the perturbation matrix in
this subspace, det$\left\vert \hbar \Omega \left\langle \psi _{n,\chi
}\right\vert F_{x}\left\vert \psi _{n+k,\chi ^{\prime }}\right\rangle
-E^{(1)}\delta _{\chi \chi ^{\prime }}\right\vert =0$, is expressed
explicitly as%
\begin{equation}
\left\vert
\begin{array}{cc}
-E^{(1)} & \frac{\sqrt{2}\hbar \Omega }{2}\eta \\
\frac{\sqrt{2}\hbar \Omega }{2}\eta ^{\ast } & -E^{(1)}%
\end{array}%
\right\vert =0,  \label{SQ1}
\end{equation}%
where $\eta =\left\langle \psi _{n}^{+1}\right. \left\vert \psi
_{n+k}^{0}\right\rangle $. Note that since $\eta $ is generally small, we
have neglected its dependence on $n$ and $k$ for simplicity. It follows that
the first-order corrections of the eigenenergies are $E_{\pm }^{(1)}=\pm
\sqrt{2}\hbar \Omega \left\vert \eta \right\vert /2$, giving rise to%
\begin{equation}
\left\{
\begin{array}{c}
\tilde{E}_{n+k,0}=E_{n,0}+E_{+}^{(1)}=(n+k)\hbar \omega +\frac{\sqrt{2}}{2}%
\hbar \Omega \left\vert \eta \right\vert \\
\tilde{E}_{n,+1}=E_{n,+1}+E_{-}^{(1)}=(n+k)\hbar \omega -\frac{\sqrt{2}}{2}%
\hbar \Omega \left\vert \eta \right\vert \\
\tilde{E}_{n,-1}=E_{n,-1}%
\end{array}%
\right. ,
\end{equation}%
and the proper zeroth-order eigenstates are
\begin{equation}
\left\{
\begin{array}{c}
\left\vert \psi _{n,+1}^{(0)}\right\rangle =\frac{\sqrt{2}}{2}(\frac{\eta }{%
\left\vert \eta \right\vert }\left\vert \psi _{n,1}\right\rangle +\left\vert
\psi _{n+k,0}\right\rangle ) \\
\left\vert \psi _{n,0}^{(0)}\right\rangle =\frac{\sqrt{2}}{2}(-\frac{\eta }{%
\left\vert \eta \right\vert }\left\vert \psi _{n,1}\right\rangle +\left\vert
\psi _{n+k,0}\right\rangle ) \\
\left\vert \psi _{n,-1}^{(0)}\right\rangle =\left\vert \psi
_{n,-1}\right\rangle%
\end{array}%
\right. .  \label{SZ1}
\end{equation}%
With the states in Eq.~(\ref{SZ1}), it is straightforward to derive the
first-order perturbative eigenstates\ using the non-perturbation theory. The
resulting eigenstate takes the form of Eq.~(\ref{DPST}), i.e.,
\begin{eqnarray}
\left\vert \tilde{\psi}_{n,\chi }\right\rangle &=&\sum_{n^{\prime
}=1}^{\infty }\left( C_{\chi ,-1}^{n,n^{\prime }}\left\vert \psi _{n^{\prime
}}^{-1}\right\rangle \left\vert -1\right\rangle +C_{\chi ,0}^{n,n^{\prime
}}\left\vert \psi _{n^{\prime }}^{0}\right\rangle \left\vert 0\right\rangle
\right.  \notag \\
&&\left. +C_{\chi ,+1}^{n,n^{\prime }}\left\vert \psi _{n^{\prime
}}^{+1}\right\rangle \left\vert +1\right\rangle \right) ,
\end{eqnarray}%
where%
\begin{widetext}
\begin{eqnarray*}
C_{+1,+1}^{n,n^{\prime }} &=&\frac{\sqrt{2}}{2}\frac{\eta }{\left\vert \eta
\right\vert }+\frac{\hbar \Omega }{2}\frac{\left\langle \psi _{n^{\prime
}}^{+1}\right. \left\vert \psi _{n+k}^{0}\right\rangle }{(n-n^{\prime
}+k)\hbar \omega +\sqrt{2}\hbar \Omega /2+\hbar \delta -\hbar \varepsilon },
\\
C_{+1,0}^{n,n^{\prime }} &=&\frac{\sqrt{2}}{2}+\frac{\hbar \Omega }{2}\frac{%
\eta }{\left\vert \eta \right\vert }(1+\frac{\sqrt{2}}{2}\delta _{n^{\prime
},n+k})\frac{\left\langle \psi _{n^{\prime }}^{0}\right. \left\vert \psi
_{n}^{1}\right\rangle }{(n-n^{\prime }+k)\hbar \omega +\sqrt{2}\hbar \Omega
/2}, \\
C_{+1,-1}^{n,n^{\prime }} &=&\frac{\hbar \Omega }{2}\frac{\left\langle \psi
_{n^{\prime }}^{-1}\right. \left\vert \psi _{n+1}^{0}\right\rangle }{%
(n-n^{\prime }+k)\hbar \omega +\sqrt{2}\hbar \Omega /2-\hbar \delta -\hbar
\varepsilon }, \\
C_{0,+1}^{n,n^{\prime }} &=&-\frac{\sqrt{2}}{2}\frac{\eta }{\left\vert \eta
\right\vert }+\frac{\hbar \Omega }{2}\frac{\left\langle \psi _{n^{\prime
}}^{+1}\right. \left\vert \psi _{n+k}^{0}\right\rangle }{(n-n^{\prime
}+k)\hbar \omega -\sqrt{2}\hbar \Omega /2+\hbar \delta -\hbar \varepsilon },
\\
C_{0,0}^{n,n^{\prime }} &=&\frac{\sqrt{2}}{2}-\frac{\hbar \Omega }{2}\frac{%
\eta }{\left\vert \eta \right\vert }(1+\frac{\sqrt{2}}{2}\delta _{n^{\prime
},n+k})\frac{\left\langle \psi _{n^{\prime }}^{0}\right. \left\vert \psi
_{n}^{1}\right\rangle }{(n-n^{\prime }+k)\hbar \omega -\sqrt{2}\hbar \Omega
/2}, \\
C_{0,-1}^{n,n^{\prime }} &=&\frac{\hbar \Omega }{2}\frac{\left\langle \psi
_{n^{\prime }}^{-1}\right. \left\vert \psi _{n+1}^{0}\right\rangle }{%
(n-n^{\prime }+k)\hbar \omega -\sqrt{2}\hbar \Omega /2-\hbar \delta -\hbar
\varepsilon }, \\
C_{-1,+1}^{n,n^{\prime }} &=&0,\ C_{-1,-1}^{n,n^{\prime }}=1, \\
C_{-1,0}^{n,n^{\prime }} &=&\frac{\sqrt{2}\hbar \Omega }{2}\frac{%
\left\langle \psi _{n^{\prime }}^{0}\right. \left\vert \psi
_{n}^{-1}\right\rangle }{(n-n^{\prime })\hbar \omega +\hbar \delta +\hbar
\varepsilon }.
\end{eqnarray*}%
\end{widetext}Following exactly the same procedure, we can readily obtain
the eigenenergies and eigenstates for the case of\ $E_{n,-1}=E_{n+k,0}$.
Accurate up to first order in $\Omega $, the eigenenergies are given by%
\begin{equation}
\left\{
\begin{array}{c}
\tilde{E}_{n,+1}=E_{n,+1} \\
\tilde{E}_{n+k,0}=(n+k)\hbar \omega +\frac{\sqrt{2}}{2}\hbar \Omega
\left\vert \eta \right\vert \\
\tilde{E}_{n,-1}=(n+k)\hbar \omega -\frac{\sqrt{2}}{2}\hbar \Omega
\left\vert \eta \right\vert%
\end{array}%
\right. ,
\end{equation}%
and the coefficients $C_{\chi ,\chi ^{\prime }}^{n,n^{\prime }}$ in the
eigenstate $\left\vert \tilde{\psi}_{n,\chi }\right\rangle $ become%
\begin{widetext}
\begin{eqnarray*}
C_{+1,+1}^{n,n^{\prime }} &=&1,\ C_{+1,-1}^{n,n^{\prime }}=0, \\
C_{+1,0}^{n,n^{\prime }} &=&\frac{\sqrt{2}\hbar \Omega }{2}\frac{%
\left\langle \psi _{n^{\prime }}^{0}\right. \left\vert \psi
_{n}^{+1}\right\rangle }{(n-n^{\prime })\hbar \omega -\hbar \delta +\hbar
\varepsilon }, \\
C_{0,+1}^{n,n^{\prime }} &=&\frac{\hbar \Omega }{2}\frac{\left\langle \psi
_{n^{\prime }}^{+1}\right. \left\vert \psi _{n+k}^{0}\right\rangle }{%
(n-n^{\prime }+k)\hbar \omega +\sqrt{2}\hbar \Omega /2+\hbar \delta -\hbar
\varepsilon }, \\
C_{0,0}^{n,n^{\prime }} &=&\frac{\sqrt{2}}{2}+\frac{\hbar \Omega }{2}\frac{%
\eta }{\left\vert \eta \right\vert }(1+\frac{\sqrt{2}}{2}\delta _{n^{\prime
},n+k})\frac{\left\langle \psi _{n^{\prime }}^{0}\right. \left\vert \psi
_{n}^{-1}\right\rangle }{(n-n^{\prime }+k)\hbar \omega +\sqrt{2}\hbar \Omega
/2}, \\
C_{0,-1}^{n,n^{\prime }} &=&\frac{\hbar \Omega }{2}\frac{\left\langle \psi
_{n^{\prime }}^{-1}\right. \left\vert \psi _{n+1}^{0}\right\rangle }{%
(n-n^{\prime }+k)\hbar \omega +\sqrt{2}\hbar \Omega /2-\hbar \delta -\hbar
\varepsilon }, \\
C_{-1,+1}^{n,n^{\prime }} &=&\frac{\hbar \Omega }{2}\frac{\left\langle \psi
_{n^{\prime }}^{+1}\right. \left\vert \psi _{n+k}^{0}\right\rangle }{%
(n-n^{\prime }+k)\hbar \omega -\sqrt{2}\hbar \Omega /2+\hbar \delta -\hbar
\varepsilon }, \\
C_{-1,0}^{n,n^{\prime }} &=&\frac{\sqrt{2}}{2}+\frac{\hbar \Omega }{2}\frac{%
\eta }{\left\vert \eta \right\vert }(1+\frac{\sqrt{2}}{2}\delta _{n^{\prime
},n+k})\frac{\left\langle \psi _{n^{\prime }}^{0}\right. \left\vert \psi
_{n}^{-1}\right\rangle }{(n-n^{\prime }+k)\hbar \omega -\sqrt{2}\hbar \Omega
/2}, \\
C_{-1,-1}^{n,n^{\prime }} &=&\frac{\sqrt{2}}{2}\frac{\eta }{\left\vert \eta
\right\vert }+\frac{\hbar \Omega }{2}\frac{\left\langle \psi _{n^{\prime
}}^{-1}\right. \left\vert \psi _{n+1}^{0}\right\rangle }{(n-n^{\prime
}+k)\hbar \omega -\sqrt{2}\hbar \Omega /2-\hbar \delta -\hbar \varepsilon }.
\end{eqnarray*}%
\end{widetext}

\end{document}